\documentclass[twocolumn,twoside,slac_two]{revtex4}
\usepackage{graphicx}
\usepackage{fancyhdr}
\begin{document}



\newcommand{\comment}[1]{\par\noindent {\em\small [#1]}}
\renewcommand{\comment}[1]{}


\newcommand{\mysubsection}[1]{\subsection[#1]{\boldmath #1}}


\newcommand{\labe}[1]{\label{equ:#1}}
\newcommand{\labs}[1]{\label{sec:#1}}
\newcommand{\labf}[1]{\label{fig:#1}}
\newcommand{\labt}[1]{\label{tab:#1}}
\newcommand{\refe}[1]{\ref{equ:#1}}
\newcommand{\refs}[1]{\ref{sec:#1}}
\newcommand{\reff}[1]{\ref{fig:#1}}
\newcommand{\reft}[1]{\ref{tab:#1}}
\newcommand{\eq}[1]{(\refe{#1})}
\newcommand{\Eq}[1]{Eq.~(\refe{#1})}
\newcommand{\Eqs}[1]{Eqs.~(\refe{#1})}
\newcommand{\Eqss}[2]{Eqs.~(\refe{#1}) and (\refe{#2})}
\newcommand{\Eqsss}[3]{Eqs.~(\refe{#1}), (\refe{#2}), and (\refe{#3})}
\newcommand{\Figure}[1]{Figure~\reff{#1}}
\newcommand{\Fig}[1]{Fig.~\reff{#1}}
\newcommand{\Figs}[1]{Figs.~\reff{#1}}
\newcommand{\Figss}[2]{Figs.~\reff{#1} and \reff{#2}}
\newcommand{\Figuress}[2]{Figures~\reff{#1} and \reff{#2}}
\newcommand{\Figsss}[3]{Figs.~\reff{#1}, \reff{#2}, and \reff{#3}}
\newcommand{\Section}[1]{Section~\refs{#1}}
\newcommand{\Table}[1]{Table~\reft{#1}}
\newcommand{\Tab}[1]{Table~\reft{#1}}
\newcommand{\Tabs}[1]{Tables~\reft{#1}}
\newcommand{\Tabss}[2]{Tables~\reft{#1} and \reft{#2}}
\newcommand{\Tabsss}[3]{Tables~\reft{#1}, \reft{#2}, and \reft{#3}}
\newcommand{\Sec}[1]{Sect.~\refs{#1}}
\newcommand{\Secs}[1]{Sects.~\refs{#1}}
\newcommand{\Secss}[2]{Sects.~\refs{#1} and \refs{#2}}
\newcommand{\Secsss}[3]{Sects.~\refs{#1}, \refs{#2}, and \refs{#3}}
\newcommand{\Chapter}[1]{Chapter~\refs{#1}}
\newcommand{\Chapters}[1]{Chapters~\refs{#1}}
\newcommand{\Chapterss}[2]{Chapters~\refs{#1} and \refs{#2}}


\newcommand{\unit}[1]{\ensuremath{\rm\,#1}}
\newcommand{\keV}{\unit{keV}}
\newcommand{\keVc}{\unit{keV/{\it c}}}
\newcommand{\keVcc}{\unit{keV/{\it c}^2}}
\newcommand{\MeV}{\unit{MeV}}
\newcommand{\MeVc}{\unit{MeV\!/\!{\it c}}}
\newcommand{\MeVcc}{\unit{MeV\!/\!{\it c}^2}}
\newcommand{\GeV}{\unit{GeV}}
\newcommand{\GeVc}{\unit{GeV\!/\!{\it c}}}
\newcommand{\GeVcc}{\unit{GeV\!/\!{\it c}^2}}
\newcommand{\TeV}{\unit{TeV}}
\newcommand{\invfb}{\unit{fb^{-1}}}
\newcommand{\cm}{\unit{cm}}
\newcommand{\mm}{\unit{mm}}
\newcommand{\ps}{\unit{ps}}
\newcommand{\invps}{\unit{ps^{-1}}}
\newcommand{\fs}{\unit{fs}}
\newcommand{\microns}{\unit{\mu m}}
\newcommand{\mrad}{\unit{mrad}}


\newcommand{\IP}{\ensuremath{\rm IP}}
\newcommand{\signif}[1]{\ensuremath{\rm #1/\sigma_{#1}}}
\newcommand{\signifm}[1]{\ensuremath{#1/\sigma_{#1}}}
\newcommand{\pT}{\ensuremath{p_{\rm T}}}
\newcommand{\pTmin}{\ensuremath{\pT^{\rm min}}}
\newcommand{\ET}{\ensuremath{E_{\rm T}}}
\newcommand{\DLL}[2]{\ensuremath{\rm \Delta  \ln{\cal L}_{\particle{#1}\particle{#2}}}}
\newcommand{\BR}[1]{\ensuremath{\rm BR(#1)}}

\newcommand{\dms}{\ensuremath{\Delta m_{\rm s}}}
\newcommand{\dmd}{\ensuremath{\Delta m_{\rm d}}}
\newcommand{\DG}{\ensuremath{\Delta\Gamma}}
\newcommand{\DGs}{\ensuremath{\Delta\Gamma_{\rm s}}}
\newcommand{\Gs}{\ensuremath{\Gamma_{\rm s}}}
\newcommand{\Gd}{\ensuremath{\Gamma_{\rm d}}}
\newcommand{\DGsGs}{\ensuremath{\Delta\Gamma_{\rm s}/\Gamma_{\rm s}}}
\newcommand{\Dm}{\mbox{$\Delta m $}}
\newcommand{\ACP}{\ensuremath{{\cal A}^{\rm CP}}}
\newcommand{\Adir}{\ensuremath{{\cal A}^{\rm dir}}}
\newcommand{\Amix}{\ensuremath{{\cal A}^{\rm mix}}}
\newcommand{\ADelta}{\ensuremath{{\cal A}^\Delta}}
\newcommand{\phid}{\ensuremath{\phi_{\rm d}}}
\newcommand{\sinphid}{\ensuremath{\sin\!\phid}}
\newcommand{\phis}{\ensuremath{\phi_{\rm s}}}
\newcommand{\sinphis}{\ensuremath{\sin\!\phis}}
\newcommand{\db}{\ensuremath{\delta_B}}
\newcommand{\rb}{\ensuremath{r_B}}
\newcommand{\rDKpi}{\ensuremath{r_D^{K\pi}}}
\newcommand{\dDKpi}{\ensuremath{\delta_D^{K\pi}}}
\newcommand{\rDKthreepi}{\ensuremath{r_D^{K3\pi}}}
\newcommand{\dDKthreepi}{\ensuremath{\delta_D^{K3\pi}}}


\newcommand{\BAR}[1]{\overline{#1}}

\newcommand{\particle}[1]{{\ensuremath{\rm #1}}}

\newcommand{\pp}{\particle{pp}}
\newcommand{\ppbar}{\particle{p\BAR{p}}}

\newcommand{\cc}{\particle{c\BAR{c}}}
\renewcommand{\b}{\particle{b}}
\newcommand{\bbar}{\particle{\BAR{b}}}
\newcommand{\bb}{\particle{b\BAR{b}}}

\newcommand{\B}{\particle{B}}
\newcommand{\Bd}{\particle{B^0}}
\newcommand{\Bs}{\particle{B^0_s}}
\newcommand{\Bds}{\particle{B^0_{(s)}}}
\newcommand{\Bu}{\particle{B^+}}
\newcommand{\Bc}{\particle{B^+_c}}
\newcommand{\Lb}{\particle{\Lambda_b}}

\newcommand{\Bbar}{\particle{\BAR{B}}}
\newcommand{\Bdbar}{\particle{\BAR{B}{^0}}}
\newcommand{\Bsbar}{\particle{\BAR{B}{^0_s}}}
\newcommand{\Bdsbar}{\particle{\BAR{B}{^0}_{(s)}}}
\newcommand{\Bubar}{\particle{B^-}}
\newcommand{\Bcbar}{\particle{B^-_c}}
\newcommand{\Lbbar}{\particle{\BAR{\Lambda}_b}}

\newcommand{\Ds}{\particle{D_s}}
\newcommand{\Dsm}{\particle{D_s^-}}
\newcommand{\KKpim}{\particle{K^+K^-\pi^-}}
\newcommand{\Dsp}{\particle{D_s^+}}
\newcommand{\Dsmp}{\particle{D_s^{\mp}}}

\newcommand{\Dz}{\particle{D^0}}
\newcommand{\Dzbar}{\particle{\BAR{D}{^0}}}
\newcommand{\DzCP}{\particle{D^0_{CP}}}

\newcommand{\Jpsi}{\particle{J\!/\!\psi}}
\newcommand{\Jmm}{\particle{\Jpsi(\mu\mu)}}
\newcommand{\Jee}{\particle{\Jpsi(ee)}}

\newcommand{\KS}{\particle{K^0_S}}
\newcommand{\Kst}{\particle{K^{*0}}}
\newcommand{\Km}{\particle{K^-}}
\newcommand{\Kp}{\particle{K^+}}
\newcommand{\pim}{\particle{\pi^-}}
\newcommand{\pip}{\particle{\pi^+}}


\newcommand{\decay}[2]{\particle{#1\!\to #2}}

\newcommand{\KSpipi}{\decay{K^0_S}{\pi^+\pi^-}}
\newcommand{\Jpsiee}{\decay{\Jpsi}{e^+e^-}}
\newcommand{\Jpsimm}{\decay{\Jpsi}{\mu^+\mu^-}}
\newcommand{\Jpsill}{\decay{\Jpsi}{\ell^+\ell^-}}

\newcommand{\DsKKpi}{\decay{\Dsm}{K^+K^-\pi^-}}

\newcommand{\Bdpipi}{\decay{\Bd}{\pi^+\pi^-}}            
\newcommand{\BdKpi}{\decay{\Bd}{K^+\pi^-}}               
\newcommand{\BspiK}{\decay{\Bs}{\pi^+K^-}}               
\newcommand{\BsKK}{\decay{\Bs}{K^+K^-}}                  
\newcommand{\Bhh}{\decay{\Bds}{h^+h^-}}                  
\newcommand{\BsDspi}{\decay{\Bs}{\Dsm\pi^+}}             
\newcommand{\BsDsK}{\decay{\Bs}{\Dsmp K^{\pm}}}          
\newcommand{\BsDsmKp}{\decay{\Bs}{\Dsm K^+}}             
\newcommand{\BsDspKm}{\decay{\Bs}{\Dsp K^-}}             
\newcommand{\BsDsh}{\decay{\Bs}{\Dsm h^+}}               
\newcommand{\BdJmmKS}{\decay{\Bd}{\Jmm\KS}}              
\newcommand{\BdJeeKS}{\decay{\Bd}{\Jee\KS}}              
\newcommand{\BdJKS}{\decay{\Bd}{\Jpsi\KS}}               
\newcommand{\BdJmmKst}{\decay{\Bd}{\Jmm\Kst}}            
\newcommand{\BdJeeKst}{\decay{\Bd}{\Jee\Kst}}            
\newcommand{\BdJKst}{\decay{\Bd}{\Jpsi\Kst}}             
\newcommand{\BuJmmK}{\decay{\Bu}{\Jmm K^+}}              
\newcommand{\BuJeeK}{\decay{\Bu}{\Jee K^+}}              
\newcommand{\BuJK}{\decay{\Bu}{\Jpsi K^+}}               
\newcommand{\BsJmmphi}{\decay{\Bs}{\Jmm\phi}}            
\newcommand{\BsJeephi}{\decay{\Bs}{\Jee\phi}}            
\newcommand{\BsJphi}{\decay{\Bs}{\Jpsi\phi}}             

\newcommand{\Bsmm}{\decay{\Bs}{\mu^+\mu^-}}              
\newcommand{\BdmmKst}{\decay{\Bd}{\mu^+\mu^-\Kst}}       
\newcommand{\BdeeKst}{\decay{\Bd}{e^+e^-\Kst}}           
\newcommand{\BdllKst}{\decay{\Bd}{\ell^+\ell^-\Kst}}     
\newcommand{\BdphiKS}{\decay{\Bd}{\phi\KS}}              
\newcommand{\Bsphiphi}{\decay{\Bs}{\phi\phi}}            
\newcommand{\BdDstpi}{\decay{\Bd}{D^{*-}\pi^+}}          
\newcommand{\BdDstpiincl}{\decay{\Bd}{D^{*-}(incl)\pi^+}}
\newcommand{\BdDKst}{\decay{\Bd}{\Dz\Kst}}               
\newcommand{\BdDbarKst}{\decay{\Bd}{\BAR{D}{^0}\Kst}}    
\newcommand{\BdDCPKst}{\decay{\Bd}{\DzCP\Kst}}           
\newcommand{\BdDbarKpiKst}{\decay{\Bd}{\Dzbar(K\pi)\Kst}}
\newcommand{\BdDbarKKKst}{\decay{\Bd}{\Dzbar(KK)\Kst}}   
\newcommand{\BdDbarpipiKst}{\decay{\Bd}{\Dzbar(\pi\pi)\Kst}}   
\newcommand{\BdDCPKKKst}{\decay{\Bd}{\DzCP(KK)\Kst}}   
\newcommand{\BdKstrho}{\decay{\Bd}{K^{*0}\rho/\omega}}   
\newcommand{\Bsetacphi}{\decay{\Bs}{\eta_c\phi}}         
\newcommand{\BsetacKKphi}{\decay{\Bs}{\eta_c(4K)\phi}}   
\newcommand{\BsetacpiKphi}{\decay{\Bs}{\eta_c(2pi2K)\phi}}  
\newcommand{\Bsetacpipiphi}{\decay{\Bs}{\eta_c(4pi)\phi}}
\newcommand{\BsJmmeta}{\decay{\Bs}{\Jmm\eta}}            
\newcommand{\BsJeeeta}{\decay{\Bs}{\Jee\eta}}            
\newcommand{\BsJeta}{\decay{\Bs}{\Jpsi\eta}}             
\newcommand{\BsJetap}{\decay{\Bs}{\Jpsi\eta'}}           
\newcommand{\BdKstgam}{\decay{\Bd}{K^{*0}\gamma}}        
\newcommand{\BdKstpin}{\decay{\Bd}{K^{*0}\pi^0}}         
\newcommand{\Bsphigam}{\decay{\Bs}{\phi\gamma}}          
\newcommand{\Bsphipin}{\decay{\Bs}{\phi\pi^0}}           
\newcommand{\Bdpipipi}{\decay{\Bd}{\pi^+\pi^-\pi^0}}     
\newcommand{\Bdrhopi}{\decay{\Bd}{\rho\pi}}              

\newcommand{\BcJmmpi}{\decay{\Bc}{\Jmm\pi^+}}            
\newcommand{\BcJpi}{\decay{\Bc}{\Jpsi\pi^+}}             
\newcommand{\Bsmumu}{\decay{\Bs}{\mu^+\mu^-}}            

\title{Flavor Physics and CP Violation at LHC}

%

\author{A. Schopper}
\affiliation{CERN, PH-Department, 1211 Geneva 23, Switzerland}

\begin{abstract}
Flavor Physics at LHC will contribute significantly to the search
for New Physics via precise and complementary measurements of CKM
angles and the study of loop decays. Here we present the expected
experimental sensitivity and physics performance of the LHC
experiments that will to B-physics.
\end{abstract}

\maketitle

\thispagestyle{fancy}


\section{Introduction}
The \B-factories are extremely successful in constraining the
Unitarity Triangles within the Standard Model and experiments at
the Tevatron have demonstrated their \Bs-physics capability.
Starting from summer 2007, the Large Hadron Collider (LHC) at CERN
shall contribute to further improve the CKM consistency test and
look for deviations from the Standard Model rare processes. LHC
will not only give access to a new high energy frontier, but will
also act as a new generation \b-factory with large \b-quark
production rates including \Bs. New Physics can be hidden in
\B-decays, since New Physics models can introduce new particles,
dynamics and symmetries at higher energy scales with virtual
particles that appear e.g. in loop processes, such as box and
penguin diagrams. Therefore, \B-physics measurements are
complementary to direct searches and will allow to understand the
nature and flavor structure of possible New Physics.

The \B-physics program at LHC is vast. It will include a precise
measurement of \Bs-\Bsbar\ mixing via e.g. \BsDspi, \BsJphi\ and
\BsJeta, to extract \dms, \DGs\ and the weak phase \phis. Possible
effects of New Physics appearing in suppressed and rare exclusive
and inclusive \B-decays will be searched for in $\Bds \to X
\gamma$, $\Bd \to \Kst l^+l^-$, $\b \to s l^+l^-$ and \Bsmm. In
order to disentangle possible New Physics contributions, the CKM
angle $\gamma$ shall be determined precisely from tree-level only
decays like \BsDsK, \BdDKst, $\B^\pm \to \Dz K^\pm$ and be
compared with the value extracted from those decays that include
loop diagrams, like \Bdpipi and \BsKK. Measurements of other CP
phases in various channels like \BdphiKS, \Bsphiphi, $\Bd \to \rho
\pi$ and $\Bd \to \rho \rho$ will further allow to over-constrain
the Unitarity Triangles.

\section{B-physics experiments at LHC}
The LHC machine will collide protons at 14~\TeV\/  center of mass
energy with a bunch crossing rate of 40$\unit{Mhz}$. At this
energy the \b\bbar\ production cross section is huge and will be
of the order of 500~$\mu b$, producing on average \b-hadrons with
about 40\% of \Bd\&\Bdbar, 40\% of \Bu\&\Bubar, 10\% of
\Bs\&\Bsbar\ and 10\% of \b-baryons. The ratio of \b\bbar\
production cross section over inelastic cross section is of the
order of 0.6\%, which requires top-performing triggers to select
the useful \B-decays.

The sensitivity of the experiments that will do \B-physics at LHC
will depend on their detector acceptance for the relevant
\B-decays, their trigger performance including fully hadronic
decays, their capability in rejecting background which requires a
good mass resolution and particle identification, their decay-time
resolution for reconstructing time-dependent \Bs-decays, and their
flavor tagging capability. Three experiments at LHC intend to do
\B-physics.

The two general-purpose experiments, ATLAS and CMS, are optimized
for discovery physics and will complete most of their \B~physics
program within the first few years~\cite{atlas}\cite{cms}, when
the LHC luminosity is expected to be below
2$\times10^{33}\cm^{-2}\unit{s}^{-1}$. In the following years of
high luminosity running with order $10^{34}\cm^{-2}\unit{s}^{-1}$,
several pp~collisions will pile-up per bunch crossing, which will
limit the \B-physics studies to the search of very rare \B-decays
with clear signatures, like e.g. $\Bds\to\mu\mu$. The reach in
\B-physics will very much depend on the trigger strategy and
bandwidth allocation. \B-events will mainly be selected by high
\pT\/ single muon and dimuon triggers. At low luminosity, ATLAS
foresees a flexible trigger strategy in which both, a muon signal
and either an electromagnetic cluster in a region of interest can
be identified to select e.g. \BdKstgam, or hadronic \b-decay
products in a jet region of interest can be identified to select
e.g. \BsDspi. CMS exploits the possibility of on-line tracking
with a reduced number of hits per track at the High Level Trigger
to select exclusive \B-events like e.g. \BsDspi.

LHCb is the experiment dedicated to B-physics at the
LHC~\cite{lhcb}. The detector is a single arm forward spectrometer
covering a pseudo-rapidity range of 1.9$\le \eta \le $4.9, which
maximizes the acceptance for \B-events, since at LHC \b\bbar\
events are produced correlated in space and are forward peaked. In
order to minimize pile-up of pp~collisions per bunch crossing,
LHCb will be running at a nominal luminosity of
2$\times10^{32}\cm^{-2}\unit{s}^{-1}$, which can be tuned locally
at the LHCb interaction point by adjusting the beam focus. Since
this luminosity is expected to be available at LHC very soon after
the start-up, LHCb shall be fully efficient staring with the first
physics run. In a nominal year with $10^7\unit{s}$/year of data
taking an integrated luminosity of $2\unit{fb}^-1$/year is
expected, translating into $10^{12}$~\b\bbar~events/year. The LHCb
trigger is optimized for selecting efficiently many different
\B-decays and operates in two stages. A fully synchronized
hardware trigger based on custom electronics boards reduces the
10\unit{MHz} visible bunch crossing rate to 1\unit{MHz}, requiring
the presence of high \pT\ leptons or photons or hadrons with a
typical \pT~cut of e.g. 1.3\GeVc\ for muons. A software trigger
running on a computer farm of about 2000 CPU's is then reducing
the output rate further to 2\unit{kHz} using the full detector
information. First it selects events with high impact parameter
and high \pT~tracks on which it then provides full event
reconstruction. The final data stream will consist of typically
200\unit{Hz} exclusive B-events and 1.8\unit{kHz} of inclusive
channels that will also be used for calibration purposes and
systematic studies.

\section{Prospects for $\Bs-\Bsbar$ mixing}

\subsection{Determination of \dms from \BsDspi}
A first measurement of the \Bs\Bsbar~oscillation frequency $\dms =
17.33^{+0.42}_{-0.21} (stat.) \pm0.07(syst.)\invps$ has been
reported in this conference by CDF, following the upper and lower
bound that was announced previously by D0.

At LHC, the determination of \dms\ from \BsDspi with better than
five sigma significance is one of the first goals of LHCb. This
requires very good proper-time resolution, flavor tagging and
background discrimination. LHCb expects an annual event yield of
80'000 events with a signal over background ratio of about 3. With
its very good proper time resolution of $\sigma_\tau \sim 40
\unit{fs}$ and a tagging power for \Bs\ of $\sim 7\%$, a five
sigma significance can be reached with the statistics of one month
of data taking. The expected proper-time distribution for
simulated \BsDspi events is shown in \Fig{rate}.

ATLAS will have a $5\sigma$ observation of oscillations after 3
years of low luminosity running, whilst the expectations for CMS
are somewhat lower due to limitations in the allocated trigger
bandwidth for \BsDspi.

\begin{figure}[h]
\centering
\includegraphics[width=\columnwidth]{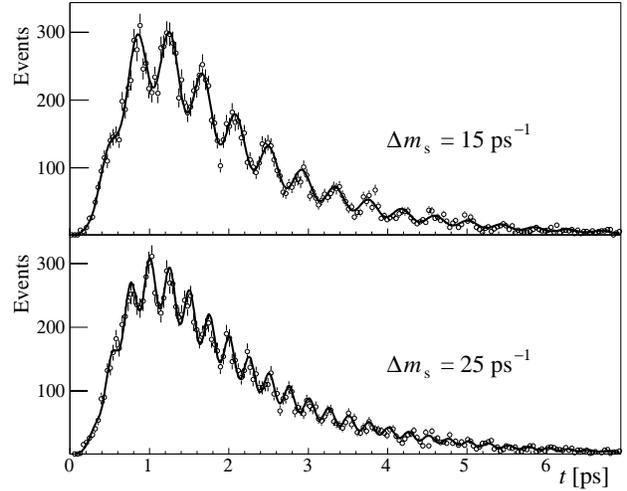}
\caption{Proper-time distribution of simulated \BsDspi\ candidates
in LHCb that have been flavour-tagged as having not oscillated,
for two different values of \dms. The data points represent one
year of data, while the curves correspond to the maximized
likelihood.} \labf{rate}
\end{figure}

\subsection{Determination of \phis\/ and \DGs from \BsJphi}
The channel \BsJphi\ is the SU(3) analogue of \BdJKS\ and can as
such be used to determine the phase \phis\ due to \Bs-\Bsbar\
oscillations. In the Standard Model, the CKM picture predicts that
this phase difference should be small,  $\phis = -2 \chi = -2 \eta
\lambda^2$, of the order of $-0.04$. The observation of a large CP
asymmetry in this channel would therefore be a striking signal for
physics beyond the Standard Model. Due to the fact that both
\Jpsi\ and \particle{\phi} are vector mesons, there are three
distinct amplitudes contributing to this decay: two CP even, and
one CP odd. Fortunately, the two CP components can be disentangled
on a statistical basis by taking into account the distribution of
the so-called transversity angle, $\theta_{\mathrm{tr}}$, defined
as the angle between the positive lepton and the  \particle{\phi}
decay plane in the \Jpsi\ rest frame (see \Fig{transversity}). The
CP-even and CP-odd components are expected to have a
non-negligible relative decay-width difference \DGsGs\ of the
order of 10\%.

\begin{figure}[h]
\centering
\includegraphics[width=\columnwidth]{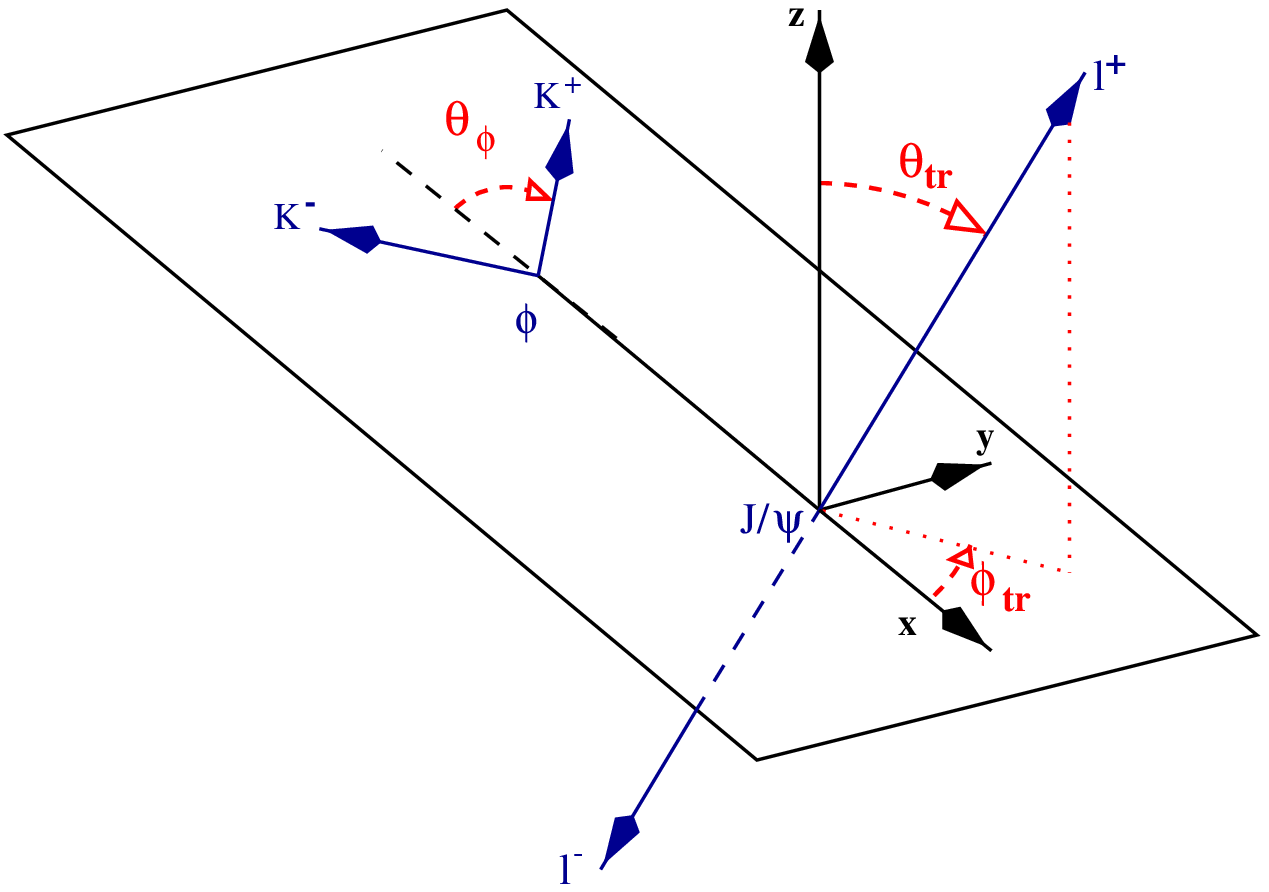}
\caption{Definition of the transversity angle
$\theta_{\mathrm{tr}}$ in the
\decay{\Bs}{\Jpsi(\ell^+\ell^-)\phi(K^+K^-)} decay.}
\labf{transversity}
\end{figure}

LHCb expects in one year of data taking to collect 125'000
\BsJphi\ decays and to obtain (for \dms = 20\invps) a precision on
sin(\phis) of 0.031 and on \DGsGs\ of about 0.011. By adding pure
CP~modes like \BsJeta\ and \BsJetap, which are expected to
contribute with about 7000 events/year, this sensitivity can be
somewhat improved. ATLAS expects a similar event rate as LHCb per
year of low luminosity running, but has a reduced sensitivity of
about 0.08 on sin(\phis).

\section{Prospects for the measurement of suppressed and rare
decays}

\subsection{Measurement of exclusive \decay{b}{s\mu\mu}}

Exclusive \decay{b}{s\mu\mu}\ decays like e.g. $\Bd \rightarrow
\Kst \mu^+\mu^-$ are suppressed decays with branching fractions of
the order of $10^{-6}$ with a clear experimental signature. The
forward-backward asymmetry is defined as
$$A_{FB}(\mbox{\^s})=(\int_0^1 dcos\theta - \int_{-1}^0 dcos\theta)
{{d\Gamma^2}\over{d\mbox{\^s} dcos\theta}}$$ where $\theta$ is the
angle between the $\mu^+$ and the $K^{*0}$ in the di-muon rest
frame, and \^s=$( m_{\mu^+\mu^-}/ m_B)^2$. The forward-backward
asymmetry is a sensitive probe of New Physics. In the Standard
Model the value of \^s for which $A_{FB}$(\^s) is zero can be
calculated with a 5\% precision. Models with non-standard values
of Wilson coefficients $C_7,C_9,C_{10}$ predict $A_{FB}$(\^s) of
opposite sign or without zero point.

LHCb will select 4400 decays per year with an expected S/B$>0.4$,
which allows a determination of the branching fractions and CP
asymmetries with a precision of a few percent. Using a toy Monte
Carlo to determine the sensitivity in the forward-backward
asymmetry measurement, including background subtraction, an
uncertainty of 0.06 on the location of \^s$_0$ is found, in 1 year
of data-taking. ATLAS will collect about 1000 $B^0_d\rightarrow
\Kst \mu^+\mu^-$ decays per year of low luminosity running, with
an expected S/B$>1$.

Other \decay{b}{s\mu\mu} decays like \decay{\Lb}{\Lambda
\mu^+\mu^-} are being investigated. The expected forward-backward
asymmetry after 3 years of low luminosity data taking by ATLAS is
shown in \Fig{afb} and compared with the expected asymmetries from
the Standard Model and from the Minimal Supersymmetric Standard
Model.

\begin{figure}[h]
\centering
\includegraphics[width=\columnwidth]{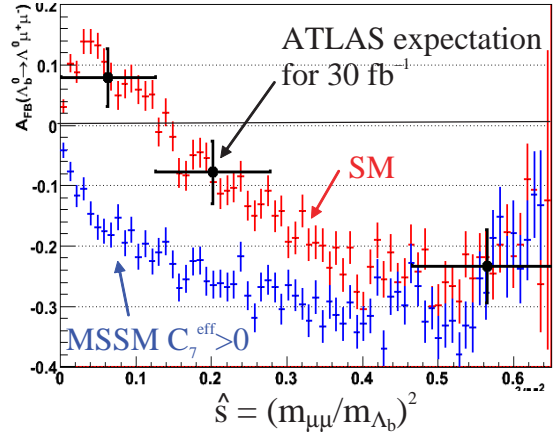}
\caption{Expected forward-backward asymmetry for
\decay{\Lb}{\Lambda \mu^+\mu^-} after 3 years of low luminosity
data taking by ATLAS (black data points), to be compared with the
expected asymmetries from the Standard Model (SM) (red) and from
the Minimal Supersymmetric Standard Model (MSSM) (blue).}
\labf{afb}
\end{figure}

\subsection{Measurement of \Bsmumu}
\Bsmumu\ is a rare decay involving flavor changing neutral
currents. In the Standard Model the branching ratio is estimated
to be $BR (B_s\rightarrow \mu^+\mu^-)=(3.5\pm0.1)\times 10^{-9}$
\cite{Ali}. In various supersymmetric extensions of the Standard
model it can be enhanced by one to three orders of magnitude with
$BR \sim (tan\beta)^6$, for large $tan\beta$. The best upper limit
on the branching ratio comes at present from CDF and is $10^{-7}$
at 95\% CL.

Within the Standard Model context, LHCb expects to select 30
signal events per year, with a resolution on the \Bs~mass of
18~MeV/c$^2$. The background determination requires a huge Monte
Carlo statistics and is still under study. Following a full
detector simulation, no background events were selected in the two
samples of $10^7$ \particle{b\bar b}\ and $10^7$ \decay{b}{\mu},
\decay{b}{\mu} events that have been used so far. The background
estimations by CMS and ATLAS rely on simulation studies with
generator cuts and assuming cut factorization. With a mass
resolution of 46~MeV/c$^2$ CMS is expecting 7 signal events with
less than 1 background event per year of low luminosity running.
ATLAS will reconstruct the \Bs~mass with a resolution of
80~MeV/c$^2$ and expects 7 signal events with less than 20
background events. Both general purpose experiments also exploit
the possibility of selecting \Bsmumu\ decays during high
luminosity runs with 30 fb$^{-1}$/year at
$10^{34}\cm^{-2}\unit{s}^{-1}$.

In conclusion there are good prospects of significant measurement
in this channel, even for the Standard Model value of the
branching ratios.

\section{Prospects for the determination of $\gamma$}

\subsection{$\gamma$ from \BsDsK\ decays}
A theoretically clean way to extract $\gamma$ is to mix the two
tree diagrams, \particle{\BAR{b}\to\BAR{u}+W^+} and
\particle{\BAR{b}\to\BAR{c}+W^+}. This can be done by studying
the time-dependent rates of \Bs\ decaying into \particle{D_s^+K^-}
and \particle{D_s^-K^+} and their CP-conjugated processes. The
measurement of two time-dependent decay asymmetries from the four
decay rates \decay{\Bs}{\particle{\Ds^\pm K^\mp}}\ and $\Bsbar
\rightarrow \particle{D_s^\mp K^\pm}$ allow to extract the phase
$\gamma+\phi_s$ together with a strong phase. Assuming that
$\phi_s$ has been determined from previous measurements, $\gamma$
can be determined with little theoretical uncertainty and is
insensitive to New Physics.

The strong particle identification capability of LHCb is essential
to separate \particle{B_s\rightarrow D_sK}\ decays from the
\particle{B_s\rightarrow D_s\pi}\ background that has a $\sim$12
times larger branching fraction. \Fig{bsdskresol} shows the mass
resolution for the reconstructed signal events and the expected
background contribution. Monte Carlo studies have shown that 5400
\particle{D_s^\mp K^\pm}\ events will be collected in one year of
data taking with a S/B ratio, estimated from \particle{b\bar{b}}\
events, larger than 1. The \particle{D_s^\mp K^\pm}\ asymmetries
are shown in \Fig{dskasym}. A sensitivity of $\sigma_\gamma=14$
degrees is obtained for $\Delta m_s$=20 ps$^{-1}$.

\begin{figure}[h]
\centering
\includegraphics[width=\columnwidth]{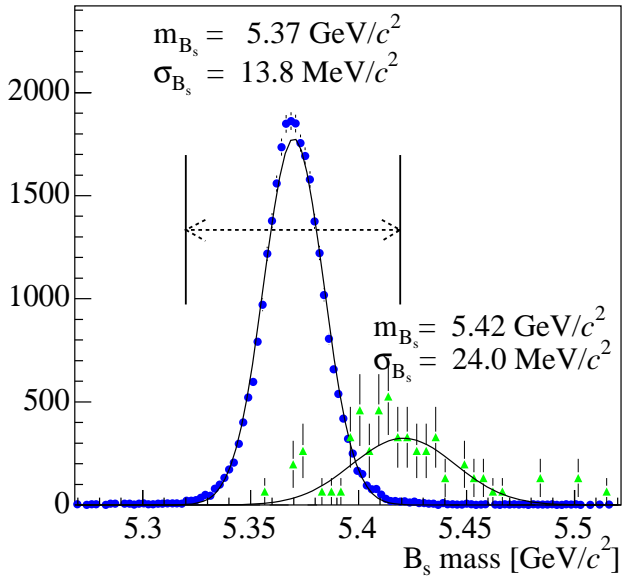}
\caption{\Bs\ mass distribution for selected \BsDsK\ candidates in
LHCb. The Gaussian fit gives a resolution of 14\MeVcc\ for the
signal. Also shown are the misidentified \BsDspi\ background
events. Their mass is shifted up due to the misidentification of
the bachelor
\particle{\pi} as a \particle{K}.}
\labf{bsdskresol}
\end{figure}

\begin{figure}[h]
\centering
\includegraphics[width=\columnwidth]{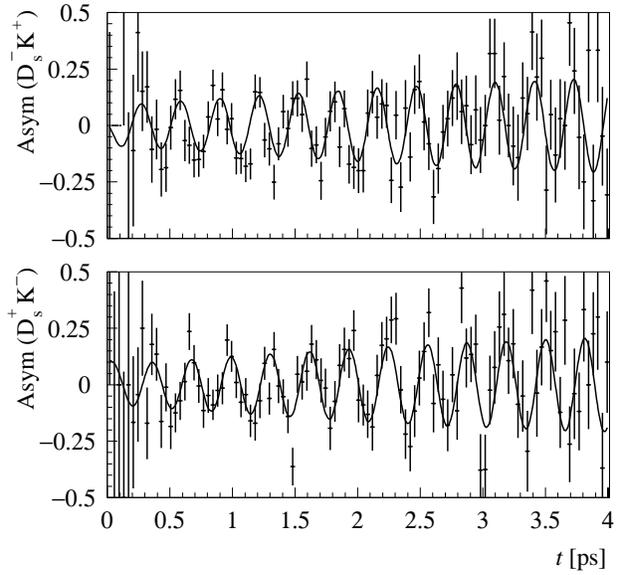}
\caption{Time-dependent \Bs\Bsbar~asymmetries of simulated
\particle{D_s^-K^+}\ (top) and \particle{D_s^+K^-} (bottom)
candidates in LHCb, for $\Delta m_s=20$ ps$^{-1}$. The errors
correspond to 5 years of data-taking.} \labf{dskasym}
\end{figure}

\subsection{$\gamma$ from \BdDKst\ decays}
The simultaneous measurement of the rates for the decays
\decay{\Bd}{\Dzbar(K^+\pi^-)\Kst}, \decay{\Bd}{\DzCP(K^+K^-)\Kst},
\decay{\Bd}{\Dz(\pi^+K^-)\Kst} and their CP conjugates, where
\decay{\Kst}{K^+\pi^-}, allows the CKM angle $\gamma$ to be
extracted, without the need of flavor tagging or proper-time
determination.

The method described in~\cite{Dunietz} is based on the measurement
of six time-integrated decay rates for \particle{B^0\rightarrow
D^0 K^{*0},\bar{D^0}K^{*0},D_{CP}K^{*0}}\ and their CP conjugates.
The decays are self-tagged through \particle{K^{*0}\rightarrow
K^+\pi^-}\, while the CP self-conjugate states \DzCP can be
reconstructed in \particle{K^+K^-}\ and \pip\pim\ modes. This
method makes use of two color-suppressed diagrams that are
interfering via \Dz~mixing, with an expected amplitude ration
$r=|A(\Bd\rightarrow \Dz \Kst)| / |A(\Bd\rightarrow \Dzbar
\Kst)|\sim 0.4$.

LHCb expects to collect 3400 \particle{\Bd\rightarrow \Dzbar\Kst},
500 \particle{\Bd\rightarrow \Dz \Kst}\ and 600
\particle{\Bd\rightarrow \DzCP\Kst}\ signal events per year of
data taking, which leads to a sensitivity for $\gamma$ of
$\sigma_\gamma \sim 8$ degrees.

\subsection{$\gamma$ from \decay{B^\pm}{\Dz K^\pm} decays}
Another approach to measure $\gamma$, closely corresponding to the
method suggested in~\cite{ads}, exploits the interference between
favored and doubly Cabibbo suppressed decays of D~mesons decaying
to states such as K$\pi$ and K$\pi\pi\pi$.

A \Bubar\ may decay color allowed into \Dz\Km\ or color suppressed
into \Dzbar\Km, with the weak phase $\gamma$ and a possible strong
phase difference \db\ between the two amplitudes. The ratio of
magnitude \rb\ between the two amplitudes is small and expected to
be of the order of 0.15. The neutral D~meson decaying into
\Kp\pim\ may arise from either a favored \Dz~decay or a doubly
Cabibbo suppressed \Dzbar~decay. The ratio of magnitude between
the two D~decay amplitudes \rDKpi\ is experimentally determined to
be of order 0.06~\cite{listing}. Taking into account a possible
strong phase difference \dDKpi\ between the two D~decay
amplitudes, one can measure the relative rates of the four
\decay{B^\pm}{\Dz K^\pm} decays, resulting in three observables
that depend on four unknown parameters $\gamma$, \db, \dDKpi\ and
\rb, and one already known parameter \rDKpi. In order to constrain
the problem it is necessary to further include D~decays into a
different final state, such as K$\pi\pi\pi$. This adds four new
rates and two new parameters, \rDKthreepi\ and \dDKthreepi, of
which the later is again experimentally determined~\cite{listing}.
Thus there are now six observables and five unknowns, which allows
to determine $\gamma$.

This method of extracting $\gamma$ from the relevant
\decay{B^\pm}{\Dz K^\pm} decay rate asymmetries is the candidate
for LHCb's statistically most precise determination of $\gamma$,
with an expected sensitivity of $\sigma_\gamma \sim 5$~degrees.

\subsection{$\gamma$ from \Bdpipi\ and \BsKK\ decays}
Extracting information on the angle $\gamma$ from two body
charmless decays of B~mesons by making assumptions on U-spin
flavor symmetry has been suggested in~\cite{fleischer}. Both
decays, \Bdpipi\ and \BsKK\ have large penguin contributions and
are therefore sensitive to New Physics.

Measuring for both decay modes the time-dependent CP~asymmetries:
\begin{equation}
\begin{array}{rcl}
A_{CP}(\Bdpipi)(t) &=& A_{CP}^{dir,\pi\pi}cos(\Delta m_d t)
\\ &&
+A_{CP}^{mix,\pi\pi}sin(\Delta m_d t) \\
&&\\
A_{CP}(\BsKK)(t)&=& A_{CP}^{dir,KK}cos(\Delta
m_s t) \\ && +A_{CP}^{mix,KK}sin(\Delta m_s t),\\
\end{array}
\end{equation}
allows to fit the four CP asymmetry coefficients. These
coefficients depend on the hadronic parameters $d$ ($d'$) and
$\vartheta$ ($\vartheta'$) that are the magnitude and phase of the
penguin-to-tree amplitude ratio of the decay transitions for
\Bdpipi\ (\BsKK). In the limit of exact U-spin symmetry of the
strong interactions, the relations $d=d'$ and
$\vartheta=\vartheta'$ hold, and the measurements of the four
asymmetry coefficients allow the simultaneous determination of
\phid\ and $\gamma$, provided that \phis\ is determined previously
from \BsJphi. Moreover, \phid\ will be accurately known by the
\BdJKS\ measurement, thus allowing a more precise determination of
$\gamma$.

The reconstruction of \Bdpipi\ and \BsKK decays requires
$K/\pi$~separation with very good efficiency and purity. This is
achieved by the particle identification system of LHCb, as shown
in \Fig{pid}. In one year of data taking, LHCb expects to
reconstruct 26'000 \Bdpipi, 37'000 \BsKK\ and 135'000 \BdKpi\
decays, with a sensitivity for the determination of $\gamma$ of
$\sigma_{\gamma}\sim 5$~degrees.

\begin{figure}[h]
\centering
\includegraphics[width=\columnwidth]{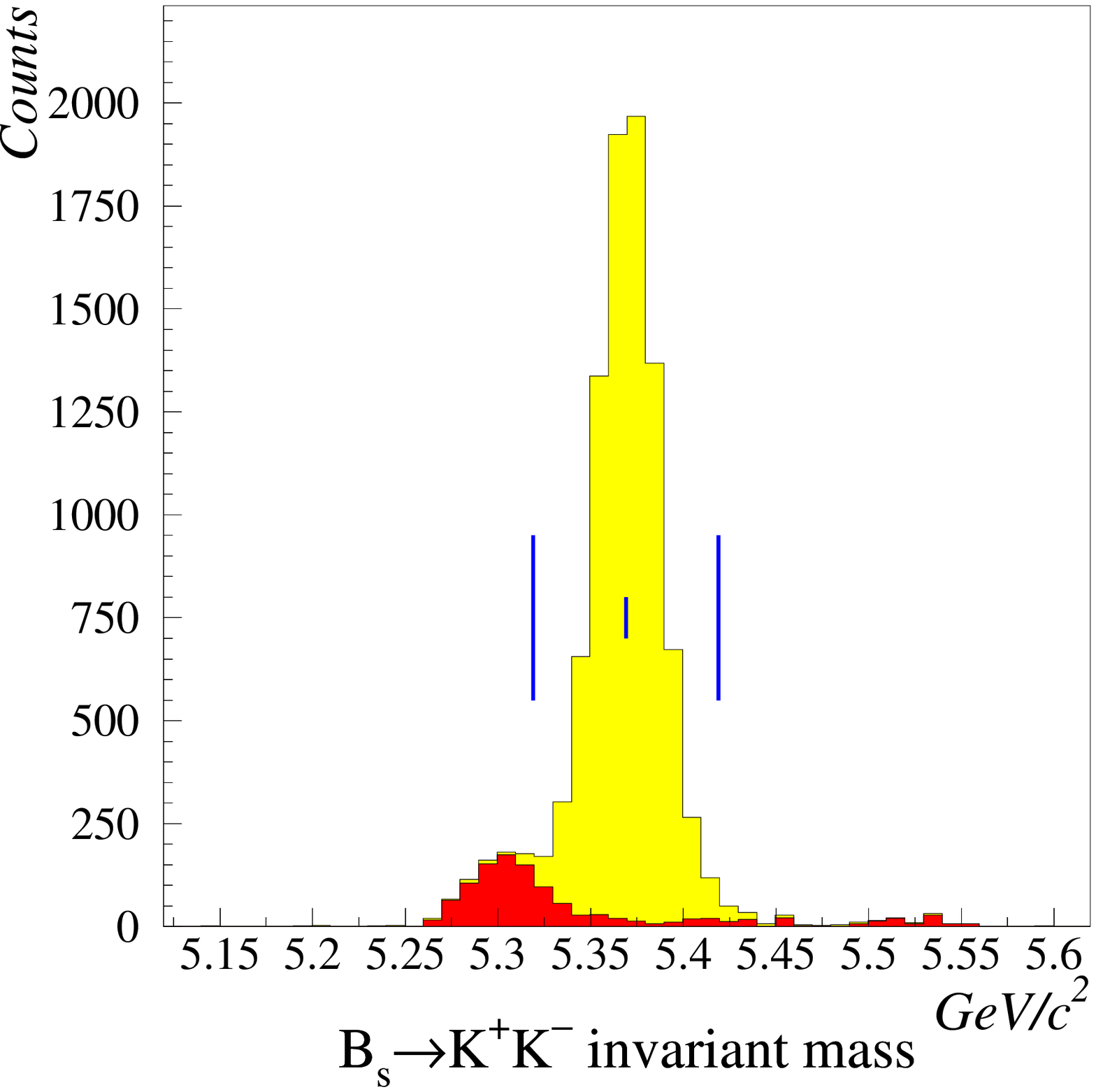}
\caption{Invariant mass distribution of selected \BsKK\
candidates. The light-shaded (yellow) histogram is the signal and
the dark (red) one represents the background from \Bdpipi, \BdKpi,
\BspiK, \decay{\Lambda_b}{pK^-} and \decay{\Lambda_b}{p\pi^-}
decays.} \labf{pid}
\end{figure}

\section{Conclusion}
In the coming years experiments at LHC will pursue an extensive
program on B-physics, complementary to the one of B-factories,
with high statistics and access to \Bs-decays. ATLAS and CMS will
contribute significantly for rare decay modes including muons, in
particular during the first years of low luminosity running. With
its dedicated flexible and robust trigger, LHCb can fully exploit
the large B-meson yields from the LHC start-up, with excellent
mass and decay-time resolution and particle identification. With
the statistics of five years of data taking one expects a
precision in the determination of $\gamma$ of $\sim 2$~degrees in
decays involving tree only, and decays involving tree and loop
diagrams. Comparing these complementary measurements will allow to
disentangle possible New Physics effects. The measurement of rare
decays and of the weak \Bs\Bsbar\ mixing phase down to a precision
better or comparable with the Standard Model predictions will also
contribute to the search for New Physics.

\begin{acknowledgments}
I would like to thank Olivier Schneider for the help in providing
many detailed information on the LHC B-physics program in general,
and to Maria Smizanska and Thomas Speer for discussion on the
ATLAS and CMS performances in particular.
\end{acknowledgments}

\bigskip 

\end{document}